\documentclass[journal,draftcls,onecolumn,12pt,twoside]{IEEEtranTCOM}
\usepackage{dcolumn}
\usepackage{fancyheadings}
\usepackage{amstext,amssymb}
\usepackage{graphicx}
\usepackage{times,color}
\usepackage[final,ulem=normalem]{changes}
\usepackage{url}
\usepackage{subfigure}
\usepackage{algorithm}
\usepackage{algorithmic}
\usepackage{latexsym}
\usepackage{amsmath}
\usepackage{flushend}

\newcommand{\comment}[1]{}

\definecolor{dgreen}{rgb}{0,0.48,0.3}

\long\def\comment#1{}

\IEEEoverridecommandlockouts

\definechangesauthor{RW}{blue}
\definechangesauthor{JW}{red}
\definechangesauthor{TC}{orange}

\begin{document}
\pagestyle{empty}
\title{LDPC Decoding with Limited-Precision\\ Soft Information in Flash Memories
\thanks{This work was presented in part at Globecom 2011 in Houston, Texas in December 2011 and at the 2012 Non-Volatile Memories Workshop at UCSD in March of 2012.  This research was supported by a gift from Inphi Corp. and UC Discovery Grant 192837.}}


\author{Jiadong Wang, Guiqiang Dong, Thomas Courtade, Hari Shankar, Tong Zhang and Richard Wesel\\
wjd@ee.ucla.edu,~dongguiqiang@gmail.com,~tacourta@ucla.edu,~hshankar@inphi.com,~tzhang@ecse.rpi.edu,~wesel@ee.ucla.edu}

\maketitle \thispagestyle{empty}
\begin{abstract}
This paper investigates the application of low-density parity-check (LDPC) codes  to Flash memories.  Multiple cell reads with distinct word-line voltages provide limited-precision soft information for the LDPC decoder.  The values of the word-line voltages (also called reference voltages) are optimized by maximizing the mutual information (MI) between the input and output of the multiple-read channel.   Constraining the maximum mutual-information (MMI) quantization to enforce a constant-ratio constraint provides a significant simplification with no noticeable loss in performance.   

Our simulation results suggest that for a well-designed LDPC code, the quantization that maximizes the mutual information will also minimize the frame error rate.  However, care must be taken to design the code to perform well in the quantized channel.  An LDPC code designed for a full-precision Gaussian channel may perform poorly in the quantized setting.  Our LDPC code designs provide an example where quantization increases the importance of absorbing sets thus changing how the LDPC code should be optimized.

Simulation results show that small increases in precision enable the LDPC code to significantly outperform a BCH code with comparable rate and block length (but without the benefit of the soft information) over a range of frame error rates.  \end{abstract}

\section{Introduction}
Flash memory can store large quantities of data in a small device that has low power consumption and no moving parts. The original NAND Flash used only two levels. This was called single-level-cell (SLC) flash because there is only one actively written charge level. Devices currently available use multiple levels and are referred to as multiple-level cell (MLC) flash. Four and eight levels are currently in use, and the number of levels will increase further \cite{LiISSCC08}\cite{TrinhISSCC08}.

Error control coding for flash memory is becoming more important in a variety of ways as the storage density increases. The increasing number of levels (and smaller distance between levels) means that variations in cell behavior from cell to cell (and over time due to wear-out) lower the signal-to-noise ratio of the read channel making a stronger error-correction code necessary. Reductions in feature size make inter-cell interference more likely, adding an equalization or interference suppression component to the read channel \cite{LeeElectron02}. Also, the wear-out effect is time\added{-}varying, introducing a need for adaptive coding to maximize the potential of the system.

Low-density parity-check (LDPC) codes are well-known for their capacity-approaching ability in the AWGN channel \cite{RichardsonDes}. LDPC codes have typically been decoded with soft reliability information while flash systems have typically provided only hard reliability information to their decoders. This paper demonstrates that at least some soft information is crucial to successfully reaping the benefits of LDPC coding in flash memory. \replaced{The paper explores}{We also explore} how much soft information is necessary to provide most of the \replaced{LDPC performance benefit}{benefits} and how flash systems could be engineered to provide the needed soft information without an unnecessary penalty in complexity or processing time.

This paper first uses pulse-amplitude modulation (PAM) with Gaussian noise to model {flash cell threshold voltage levels}, and investigates how to optimize the word-line voltages \replaced{to provide the maximum}{by maximizing the} mutual information (MMI) between {the} input and {the}  output of the equivalent read channel. After choosing the word-line voltage for each of the reads, the multiple-read channel can be represented by a discrete-alphabet channel with a known probability transition matrix and the data can be decoded with standard belief-propagation. \replaced{The paper then extends}{Then we extend} the MMI approach to \deleted{other channel models such as} \added{the} retention noise model \cite{DongUSENIX2011} \added{which more accurately models flash memories}. \replaced{The MMI}{maximum mutual information (MMI)} approach is also explored in \cite{LeeISIT05} \cite{KurkoskiIT} for the design of the message-passing decoders of LDPC codes to optimize the quantization of the \added{output of a} binary-input channel\deleted{ output}.

In \cite{DongTCAS2011}, a heuristic quantization algorithm sets the word-line voltage between any two adjacent pdfs to the value where the two pdfs have a constant ratio $ R $. \added{The constant-ratio method in \cite{DongTCAS2011} adjusts the parameter $R$  using a computationally intensive empirical method. } The MMI approach provides a natural way to select $R$ in the constant-ratio method. Simulation results show that the constant-ratio method simplifies the MMI approach with negligible performance loss for four-level cells.  

This paper also briefly explores how quantization affects LDPC code design.  LDPC codes are usually designed with the degree distribution optimized for the AWGN channel \cite{RichardsonDes}. However, our simulations show that, in the quantized setting, adjusting this ``optimal'' degree distribution to avoid small absorbing sets can significantly improve performance.

Section~\ref{background} introduces the NAND flash memory model and LDPC codes. Section~\ref{casestudy} \added{shows how to obtain word-line voltages by maximizing the MI of the equivalent read channel.  This section presents the MMI optimization approach and the constant-ratio method for SLC and MLC flash} \replaced{for}{studies} the PAM-Gaussian model and the more realistic retention noise model.  Section~\ref{results} provides simulation results demonstrating the benefits of using soft information with word-line voltages selected as described in Section~\ref{casestudy} and compares the full MMI optimization with optimization constrained using the constant\added{-}ratio method. Section~\ref{conclusion} delivers the conclusions.

\section{Background}\label{background}
This section introduces the basics of NAND flash memory and LDPC codes.

\subsection{Basics of NAND Flash Memory}\label{basic:NAND}
\begin{figure}
\centering
\includegraphics[width=0.4\textwidth]{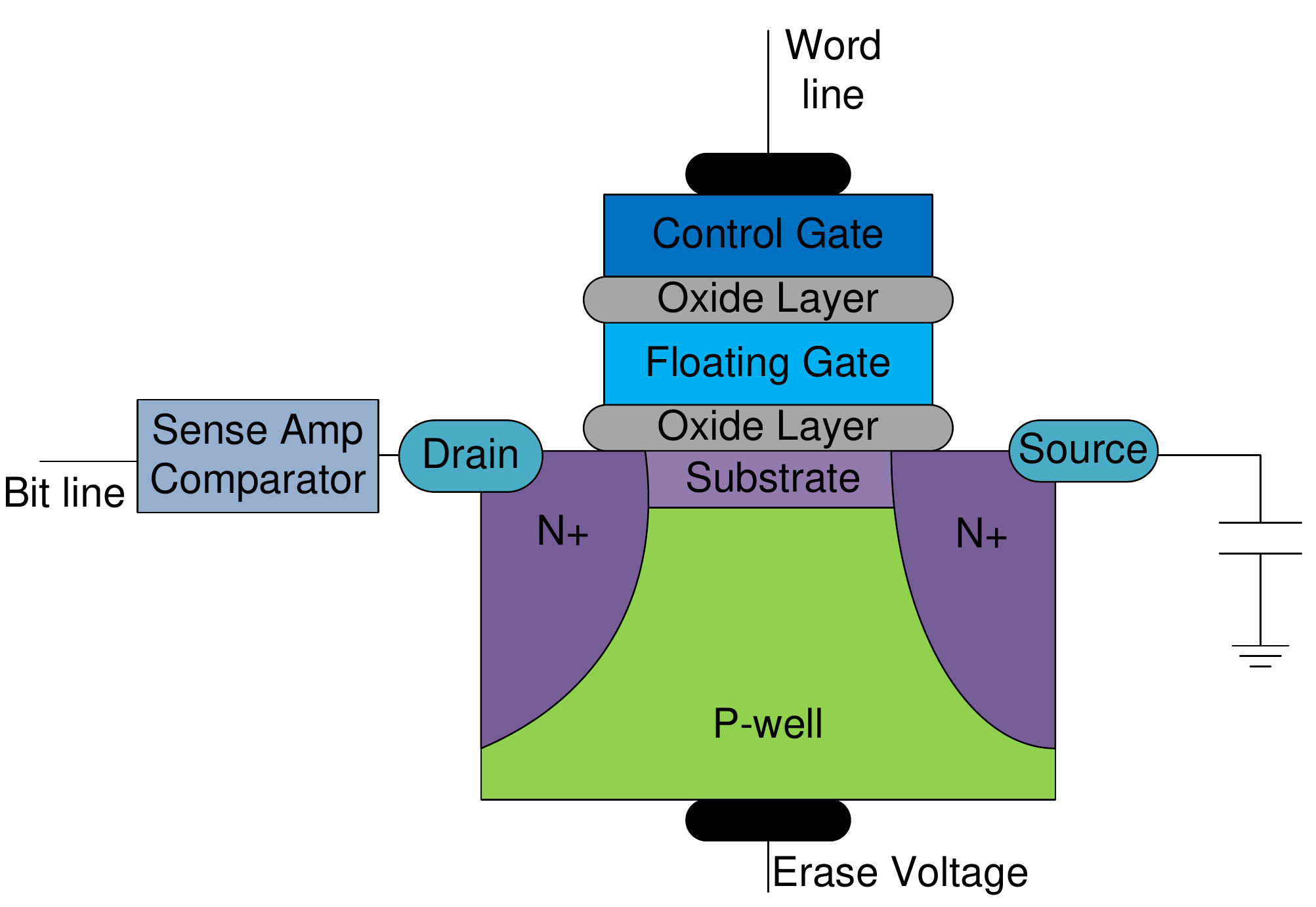}
\caption{A NAND flash memory cell.}\label{flashcell}
\end{figure}

This paper focuses on the NAND architecture for flash memory\deleted{, which is the most prevalent architecture today}. Fig. \ref{flashcell} shows the configuration of a NAND flash memory cell. Each memory cell in the NAND architecture features a transistor with a control gate and a floating gate. To store information, a charge level is written to the cell by adding a specified amount of charge to the floating gate through Fowler-Nordheim tunneling by applying a relatively large voltage to the control gate \cite{BezIEEE03}.

To read a memory cell, the charge level written to the floating gate is detected by applying a specified word-line voltage to the control gate and measuring the transistor drain current. The drain current is compared to a threshold by a sense amp comparator. If the drain current is above the comparator threshold, then the word-line voltage was sufficient to turn on the transistor, indicating that the charge written to the floating gate was insufficient to prevent the transistor from turning on. If the drain current is below the threshold, the charge added to the floating gate was sufficient to prevent the applied word-line voltage from turning on the transistor. The sense amp comparator only provides one bit of information about the charge level present in the floating gate. 

A bit error occurring at this threshold-comparison stage is called a \emph{raw bit error}  and we use the phrase \emph{channel bit error probability} to refer to the probability of a raw bit error given a specified amount of distortion in the overall process of writing to the cell, retaining the charge level over a period of time, and then reading the cell.  We refer to this overall process as the \emph{read channel}.

The word-line voltage required to turn on a particular transistor (called the threshold voltage) can vary from cell to cell for a variety of reasons. For example{, the floating gate can be overcharged during the write operation, the floating gate can lose charge due to leakage in the retention period, or the floating gate can receive extra charge when nearby cells are written \cite{MaedaISDFT09}.}  We refer to this variation of threshold voltage from its intended value as the \emph{read channel noise}.

The probability density function of the read channel noise can be modeled by a Gaussian distribution. In this paper, we \added{initially} assume an i.i.d. Gaussian threshold voltage for each level of an MLC flash memory cell \replaced{such that}{. Therefore} an $ m $-level flash cell is equivalent to an $ m $-PAM communication system with AWGN noise, except that the threshold voltage cannot be directly observed.  Rather, \added{at most} one bit of information about the threshold voltage may be obtained by each cell read.

More precise models such as the model in \cite{MaedaISDFT09} in which the lowest and highest threshold voltage distributions have a higher variance and the model in \cite{LiVLSI10} in which the lowest threshold voltage (the one associated with zero charge level) is Gaussian and the other threshold voltages have Gaussian tails but a uniform central region are sometimes used. The model in  \cite{DongUSENIX2011} is similar to \cite{LiVLSI10}, but is derived by explicitly accounting for {real dominating noise sources, such as inter-cell interference, program injection statistics, random telegraph noise and retention noise}.  \added{This paper uses the model of \cite{DongUSENIX2011} to study the MMI approach and constant ratio method in a more realistic setting to complement the analysis using a simple i.i.d. Gaussian assumption.}


\subsection{Basics of LDPC codes} \label{basic:LDPC}
LDPC codes \cite{Gallagerthesis} are linear block codes defined by sparse parity-check matrices. By optimizing the degree distribution, it is well-known that LDPC codes can approach the capacity of an AWGN channel \cite{RichardsonDes}. Several algorithms have been proposed to generate LDPC codes for a given degree distribution, such as the ACE algorithm \cite{TianTCOM04}, and the PEG algorithm \cite{PEG01}.

Designing LDPC codes with low error-floors is crucial for \replaced{the}{applications to} flash memory \added{application} since storage systems usually require block-error-rates lower than $ 10^{-15} $. This topic has generated a significant amount of recent research including \cite{richardson} \cite{JWANGICC11} \cite{WangITA2011} \cite{ivkovicIT08} \cite{NguyenITW10} \cite{HuangITA2011}.

In addition to their powerful error-correction capabilities, another appealing aspect of LDPC codes is the existence of low-complexity iterative algorithms used for decoding. These iterative decoding algorithms are called belief-propagation algorithms. Belief-propagation decoders commonly use soft reliability information about the received bits, which can greatly improve performance. Conversely, a quantization of the received information which is too coarse can degrade the performance of an LDPC code.

Traditional algebraic codes, such as BCH codes, commonly use bounded-distance decoding and can \deleted{only} correct \added{up to} a specified, fixed number of errors. Unlike these traditional codes,  {it can be difficult for LDPC codes to } guarantee a specified number of correctable errors.  However the average bit-error-rate performance can often outperform that of BCH codes in Gaussian noise.

The remainder of this paper studies how quantization during the read process affects the performance of LDPC decoding for flash memory. In the next section, we {present a general quantization approach for selecting word line voltages} for reading the flash memory cells in both the SLC and the MLC cases.

\section{Soft Information Via Multiple Cell Reads} \label{casestudy}
Because the sense amp comparator \deleted{only} provides \added{at most} one bit of information about the threshold voltage (or equivalently \added{about} the amount of charge present in the floating gate), decoders for error control codes in flash have historically \replaced{used}{relied on} hard \deleted{bit} decisions \added{on each bit} from the sense-amp comparator \added{outputs}.

 \replaced{Soft}{However, soft} information can be obtained in two ways: either by reading from the same sense amp comparator multiple times with different word line voltages (as is already done to read multi-level flash cells) or by equipping a flash cell with multiple sense amp comparators on the bit line, which is essentially equivalent to replacing the sense amp comparator (a one-bit A/D converter) with a higher-precision A/D converter.

These two approaches are not completely interchangeable. The real goal is to detect soft information about the threshold voltage. Each additional read of a single sense amp comparator can provide additional useful soft information about the threshold voltage if the word-line voltages are well-chosen. 

\replaced{In contrast,}{However,} multiple comparators may not give much additional information if the drain current vs. word-line-voltage curve (the classic I-V transistor curve) is too nonlinear. If the drain current has saturated too low or too high, the outputs from more sense-amp comparators are not useful in establishing precisely how much charge is in the floating gate. If the word-line voltage and floating gate charge level place the transistor in the linear gain region, then some valuable soft information is provided by multiple sense amp comparators.  

Our work focuses on the first technique described above in which soft information is obtained from multiple reads using the same sense-amp comparator with different word line voltages.

This section investigates the potential improvement of increasing the resolution beyond one bit and studies how best to obtain this increased resolution.  In \cite{DongTCAS2011}, the use of soft information was explored and the poor performance of uniformly spaced word-line voltages was clearly established.

This paper takes an information-theoretic perspective on optimizing the word-line voltages.  \added{Similar to other work (not in the context of flash memory) such as \cite{LeeISIT05} \cite{KurkoskiIT}, this paper seeks to quantize so as to create an effective read  channel that has the maximum mutual information (MMI).} We study quantization models with different numbers of reads for both SLC and MLC flash memory. 

In the course of our analysis, we choose the word-line voltages for each quantization \replaced{to achieve MMI}{by maximizing the MI} between the input and output of each equivalent read channel. Theoretically, this choice of word-line voltages maximizes the amount of information provided by the quantization. The next subsection studies simple PAM-Gaussian models. We first explore an example of SLC with just one additional read to provide extra soft information. Then, the section looks at the benefit of additional reads for SLC and MLC. After that, we extend the analysis to the retention noise model and study constraining the MMI approach with the constant-ratio method. Numerical results are given in Section \ref{results}.

\begin{figure}
\centering
\includegraphics[width=0.4\textwidth]{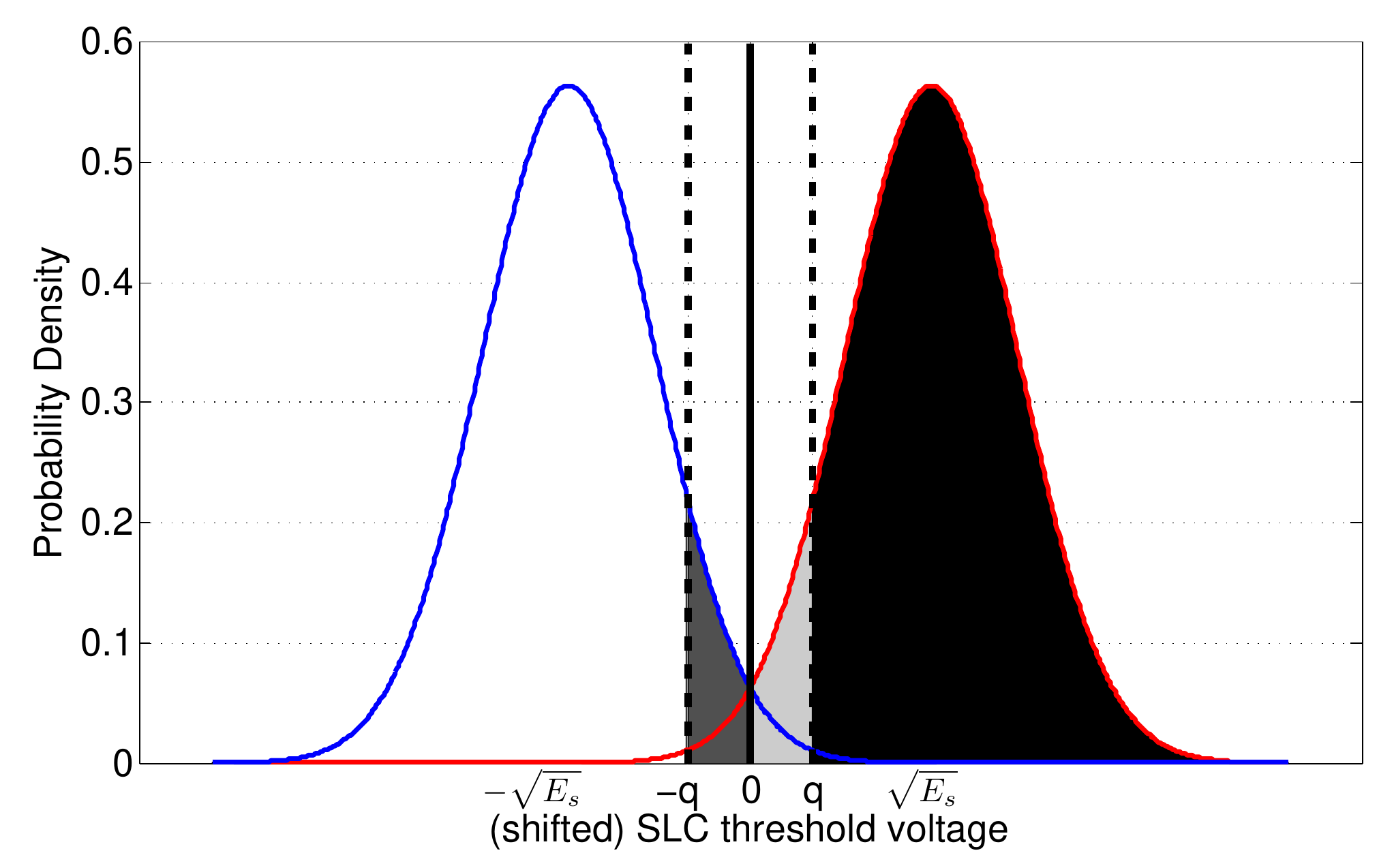}
\caption{Identically distributed Gaussian model for SLC threshold voltages.  Also shown are word-line voltages for two reads (the dashed lines) and three reads (all three lines).  The quantization regions are indicated by shading with the middle region for two reads being the union of the light and dark gray regions.}\label{quantSLC2reads}\label{quantSLC3reads}
\end{figure}

\begin{figure}
\centering
\includegraphics[width=0.4\textwidth]{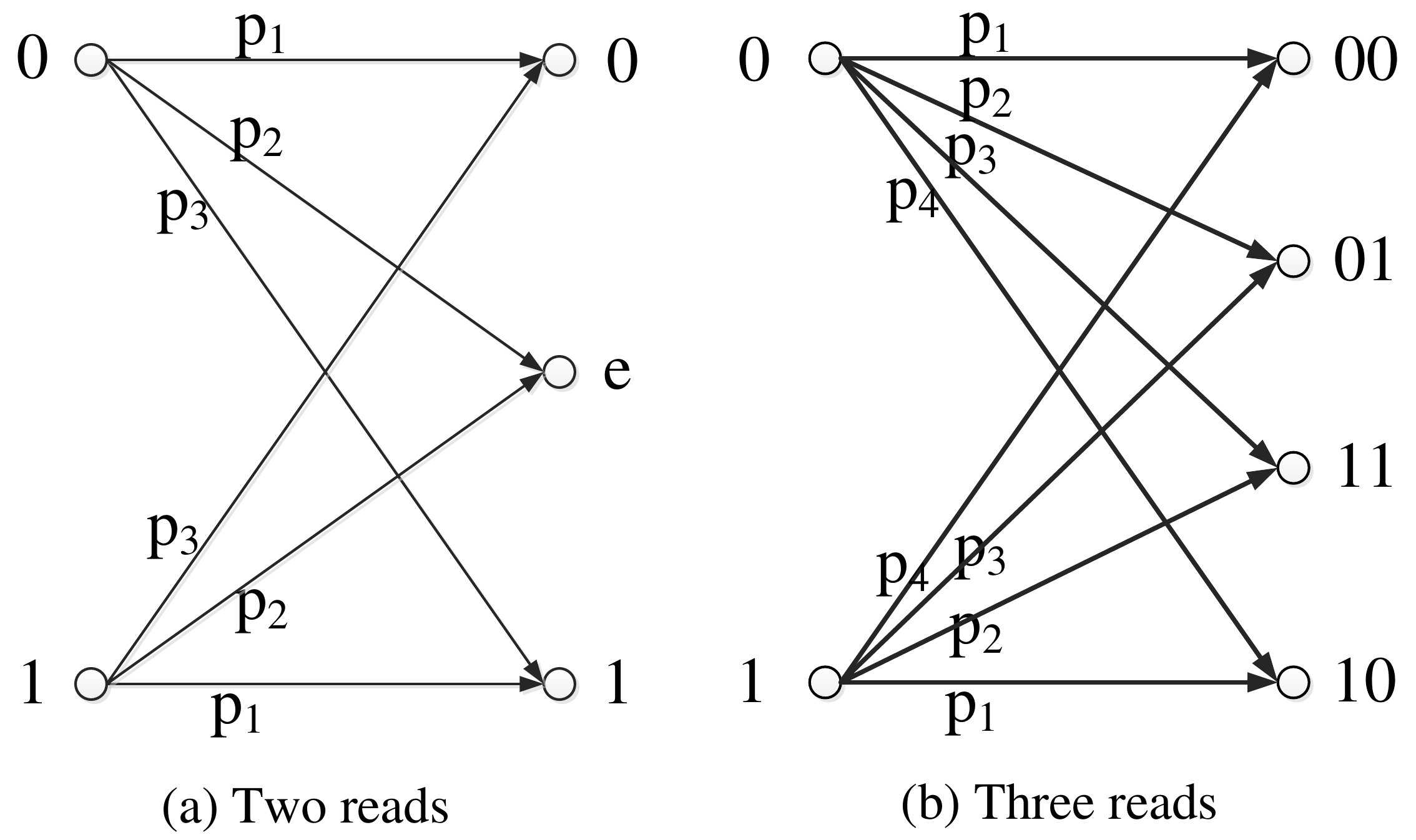}
\caption{Equivalent discrete memoryless channel models for SLC with (a) two reads and (b) three reads with distinct word-line voltages.}
\label{fig:slc2read3read}
\end{figure}

\subsection{PAM-Gaussian Model}\label{sec:gaussian}

\added{This subsection describes how to select word-line voltages to achieve MMI in the context of a simple model of the flash cell read channel as  PAM transmission with Gaussian noise.  MMI word-line voltage  selection is presented using three examples: SLC with two reads, SLC with three reads, and 4-level MLC with six reads. }

For SLC flash memory, each cell can store \replaced{one}{1} bit of information. Fig. \ref{quantSLC2reads} shows \replaced{the}{a simplistic} model of the threshold voltage distribution as a mixture of two  \added{identically distributed} Gaussian random variables.  When either a  ``0''  or  ``1'' is written to the cell, the threshold voltage is modeled as a Gaussian random variable with variance $N_0/2$ and either mean $-\sqrt{E_s}$ (for  ``1'' ) or mean $+\sqrt{E_s}$ (for  ``0'' ).

For SLC with two reads using two different word line voltages, which should be symmetric (shown as $ q $ and $ -q $ in Fig.~\ref{quantSLC2reads}), the threshold voltage is quantized according to three regions shown in Fig.~\ref{quantSLC2reads}: the white region,  the black region, and the union of the light and dark gray regions (which essentially corresponds to an erasure).  This quantization produces the effective discrete memoryless channel (DMC) model as shown in Fig.~\ref{fig:slc2read3read} (a) with input  $ X \in \{0,1\}$ and output $ Y \in \{0,e,1\} $.

Assuming $ X $ is equally likely to be 0 or 1, the MI $I(X;Y)$ between  the input $X$ and output $Y$ of the resulting DMC can be calculated \cite{CoverEOIT} as
\begin{align}
I(X;Y) &= H(Y)-H(Y|X) \notag \\
&= H\left( \frac{\text{$p_1$$+$$p_3$}}{2},p_2,\frac{\text{$p_1$$+$$p_3$}}{2}\right)-H\left(p_1,p_2,p_3\right),\label{equ:mu2reads}
\end{align}
where $H$ is the entropy function \cite{CoverEOIT} and the relevant crossover probabilities shown in Fig.~\ref{fig:slc2read3read} (a) are \mbox{$p_1=1-Q^{-}$, $p_2 = Q^{-} - Q^{+}$, and $p_3 =Q^{+}$}  with
\begin{equation}
Q^{-} = Q\left( \frac{\sqrt{E_s}-q}{\sqrt{N_0/2}} \right)\text{ and } Q^{+} = Q\left( \frac{\sqrt{E_s}+q}{\sqrt{N_0/2}}\right), \label{eq:Q-Q+}
\end{equation}
where $Q(x) = \frac{1}{\sqrt{2 \pi}} \int_x^{\infty}e^{-u^2/2}du$.

\begin{figure}
\centering
\includegraphics[width=0.4\textwidth]{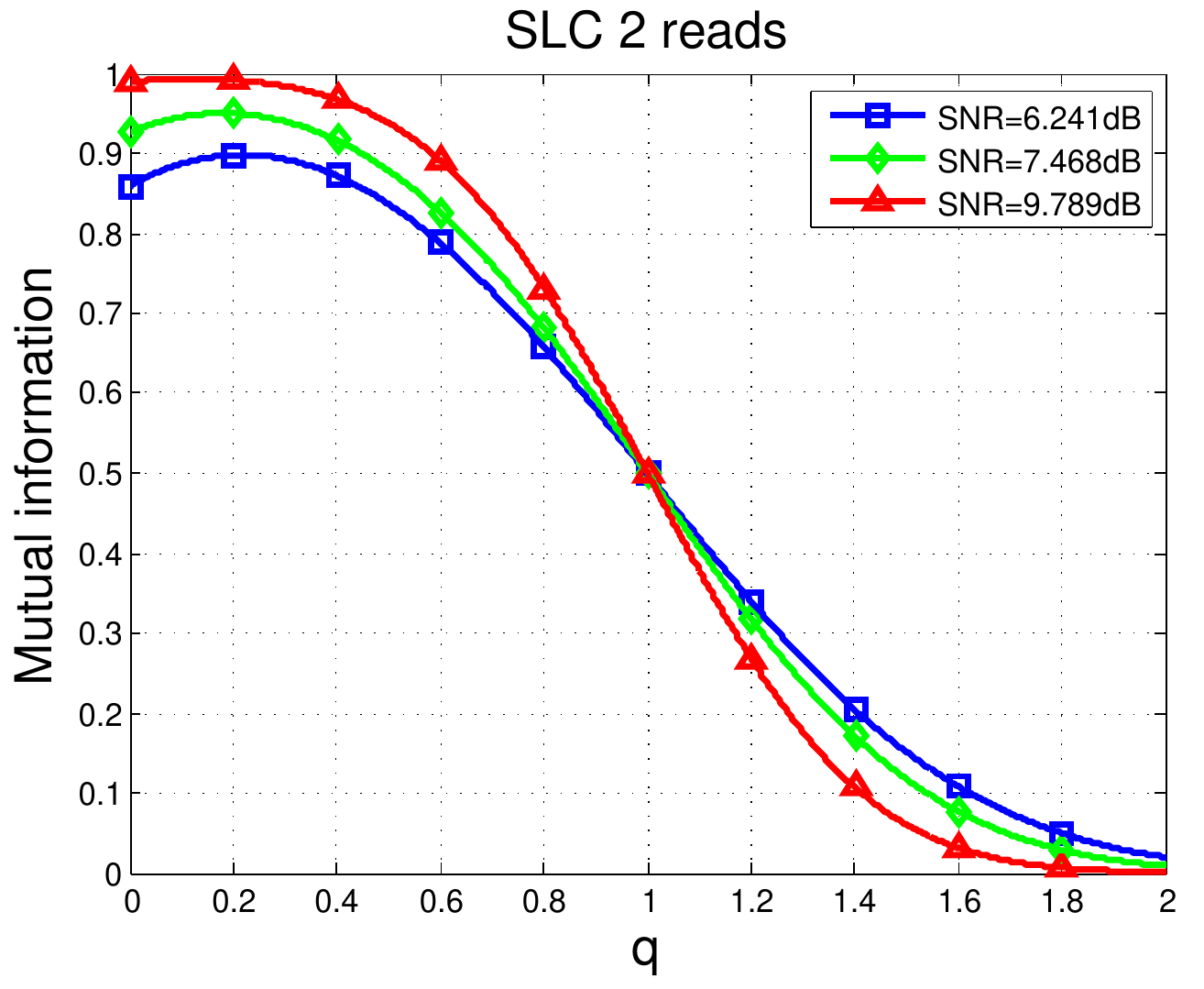}
\caption{MI vs. $q$ (for $E_s=1$) for SNR$= \frac{E_s}{N_0/2}= 6.241$ dB, 7.468 dB and 9.789 dB for SLC flash with two reads.}\label{fig:MI_vs_q_2reads}
\end{figure}

For fixed signal-to-noise ratio (SNR) $\frac{E_s}{N_0/2} $, the MI in \eqref{equ:mu2reads} is a quasi-convex function of $q$ and can be maximized numerically {to find} the parameter $q$ that yields the MMI.  Fig. \ref{fig:MI_vs_q_2reads} shows how MI varies as a function of $q$ for three values of SNR. Note that the optimum $ q^* $ is a function of SNR. For example, if  the SNR is $6.241$ dB, $ q^*=0.2188 \sqrt{E_s} $, and if the SNR is $9.789$ dB, $ q^*=0.1253 \sqrt{E_s}$.  Near the maximum MI, the variation of MI with $q$ becomes more sensitive as the channel SNR degrades.

For the simple Gaussian model of SLC with three reads for each cell,  the word-line voltages should again be symmetric (shown as $ q $, $0$, and $ -q $ in Fig.~\ref{quantSLC3reads}).  The threshold voltage is quantized according to the four differently shaded regions shown in Fig.~\ref{quantSLC3reads}.  This quantization produces the effective DMC model is shown in Fig.~\ref{fig:slc2read3read} (b) with input  $ X \in \{0,1\}$ and output $ Y \in \{00, 01, 10, 11\} $.  

Assuming $ X $ is equally likely to be 0 or 1, the MI between the input and output of this DMC can be calculated as
\begin{align}
I(X;Y) =& H(Y)-H(Y|X)\notag\\
=& H\left(\frac{p_1+p_4}{2},\frac{p_2+p_3}{2},\frac{p_3+p_2}{2},\frac{p_4+p_1}{2}\right)\notag \\
&-H(p_1,p_2,p_3,p_4),\label{equ:mu3reads}
\end{align}
where the crossover probabilities are $p_1 = 1-Q^{-}$, $p_2 = Q^{-} - Q^{0}$, $p_3 = Q^{0} - Q^{+}$, and $p_4 =Q^{+}$  with $Q^{-}$ and $Q^{+}$ as in \eqref{eq:Q-Q+} and 
\begin{equation}
 Q^{0} = Q\left( \frac{\sqrt{E_s}}{\sqrt{N_0/2}}\right).
\end{equation}

\begin{figure}
\centering
\includegraphics[width=0.4\textwidth]{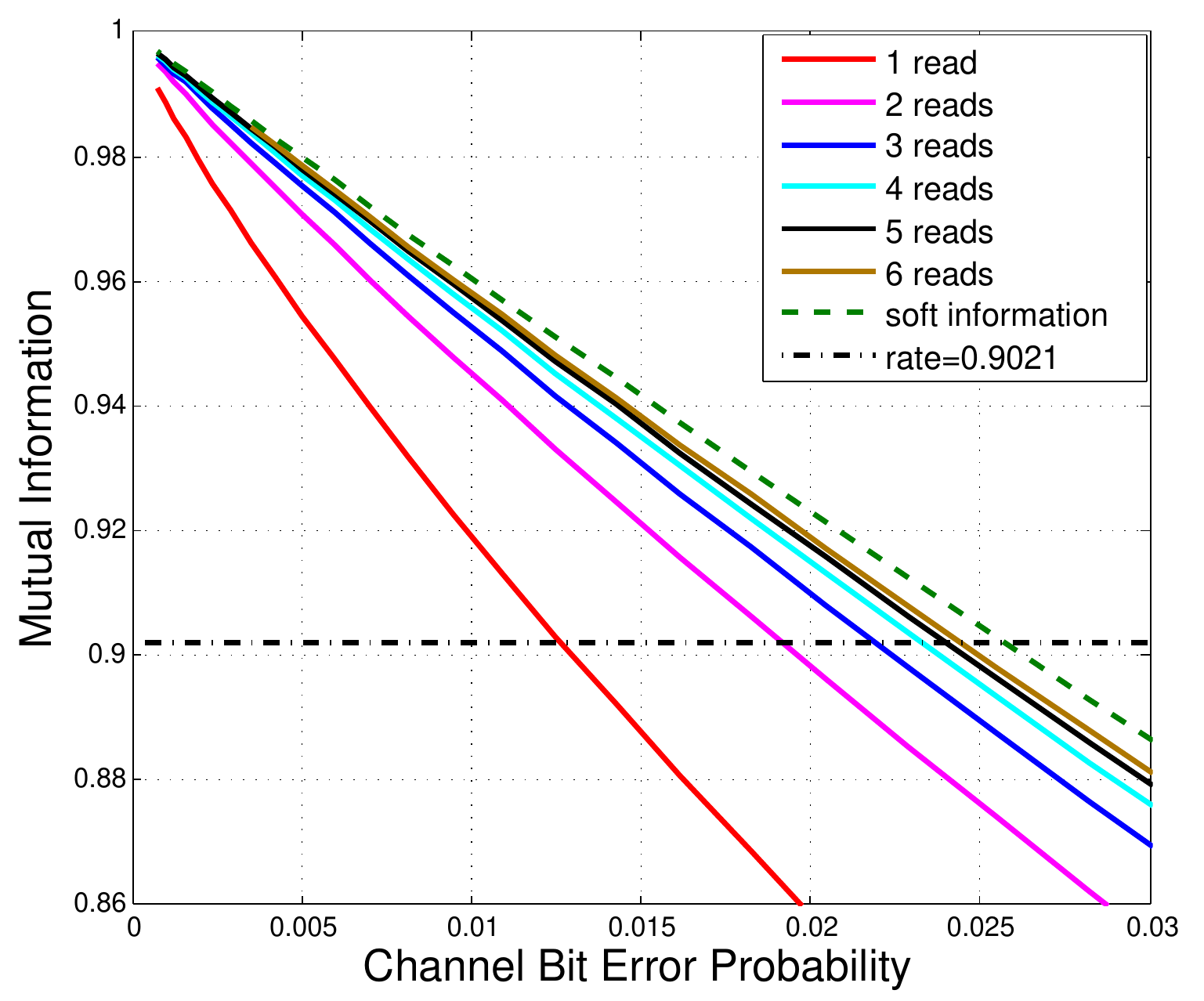}
\caption{MI provided by different quantizations for SLC.  The dashed horizontal line indicates the operating rate of our simulations.  When a MI curve is below the dashed line, the read channel with that quantization cannot possibly support the attempted rate.}\label{mutual_slc}
\end{figure}

\begin{figure}
\centering
\includegraphics[width=0.4\textwidth]{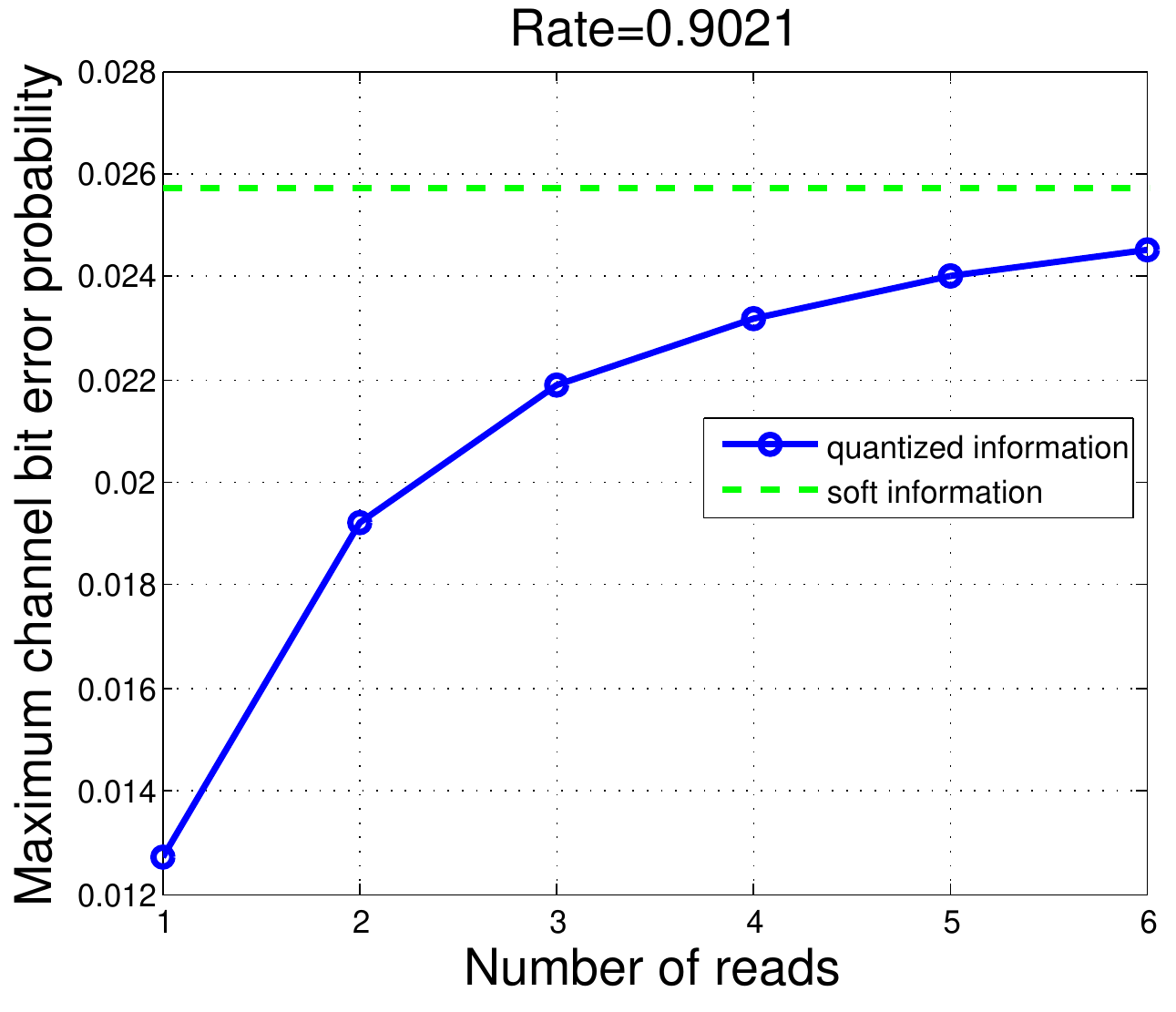}
\caption{Maximum tolerable channel bit error probability provided by different quantizations for SLC operating at rate 0.9021.}\label{fig:maxerror_vs_Nreads}
\end{figure}

We can find the MMI word-line voltages for larger numbers of reads similarly.  Fig.~\ref{mutual_slc} shows how MI increases with the number of reads for the SLC case.  The top curve shows the MI possible with full soft information (where the receiver would know the threshold voltage exactly).  The bottom curve shows the MI available with a single read, which is what is available with a typical SLC implementation.   With two reads, the MI is improved enough to close about half of the gap between the single-read MI and the MI of full soft information.  The MI with 3 reads is larger than the MI with 2 reads. This approach can be extended to however many reads are desired, but with diminishing returns as shown in Fig.~\ref{mutual_slc}.  Fig. \ref{fig:maxerror_vs_Nreads} shows how the bit error probability requirement to achieve an MI of 0.9021 increases (relaxes) as the number of reads increases.  

For 4-level MLC flash memory, each cell can store 2 bits of information. Extending the previously introduced SLC model in the natural way,  we model the MLC read channel as a 4-PAM signal with AWGN noise. To minimize the raw bit error rate, we also use the Gray labeling $ (00,01,11,10) $ for these {four} levels. Typically in 4-level MLC flash, each cell is compared to 3 word-line voltages and thus the output of the comparator has 4 possible values (i.e., four distinct quantization regions). 

\begin{figure}
\centering
\includegraphics[width=0.4\textwidth]{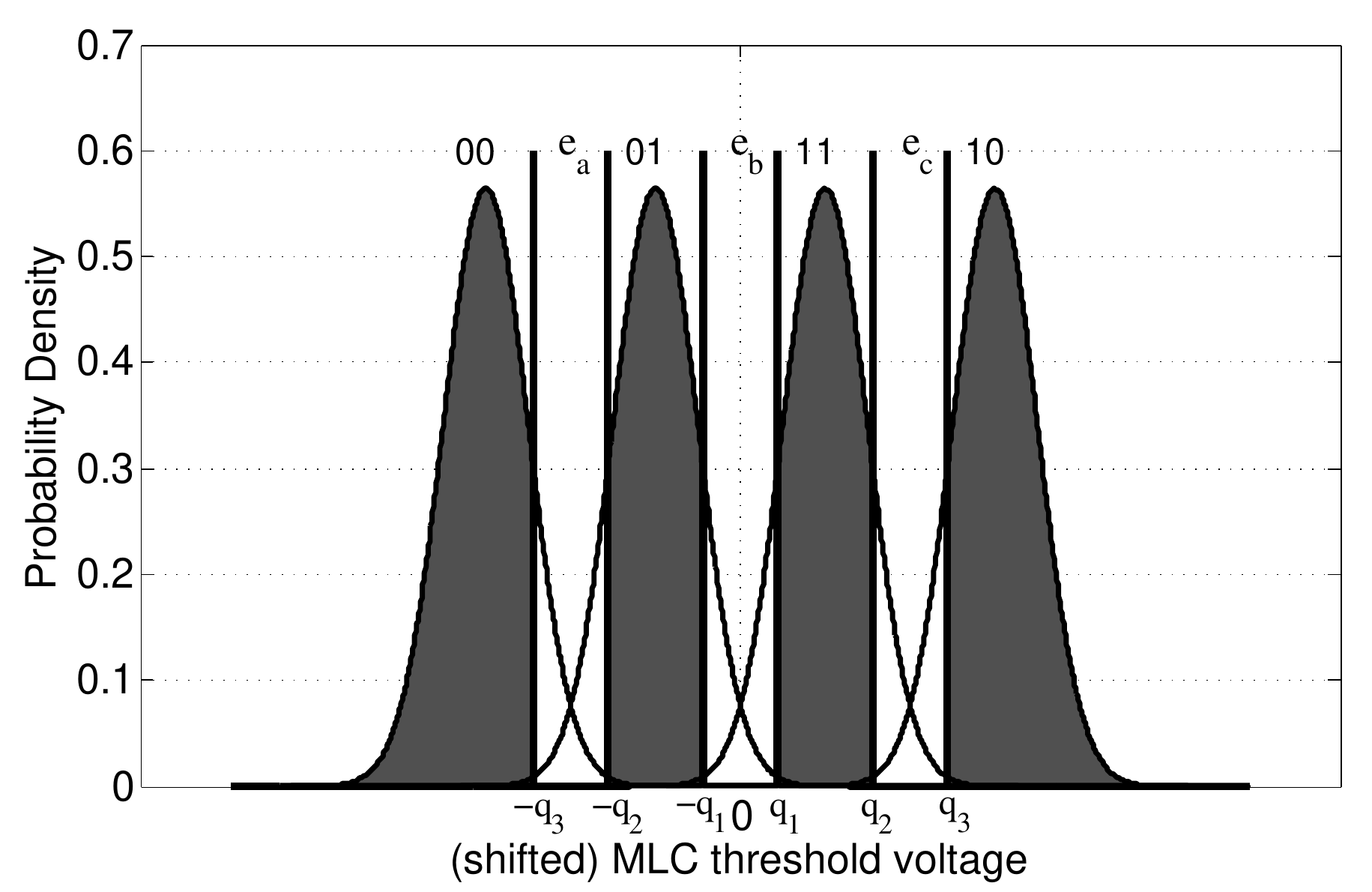}
\caption{Channel model for four-level MLC with threshold voltages modeled as Gaussians all sharing the same variance.  Quantization is shown for six reads.}\label{quant4MLC6reads}
\end{figure}

\begin{figure}
\centering
\includegraphics[width=0.4\textwidth]{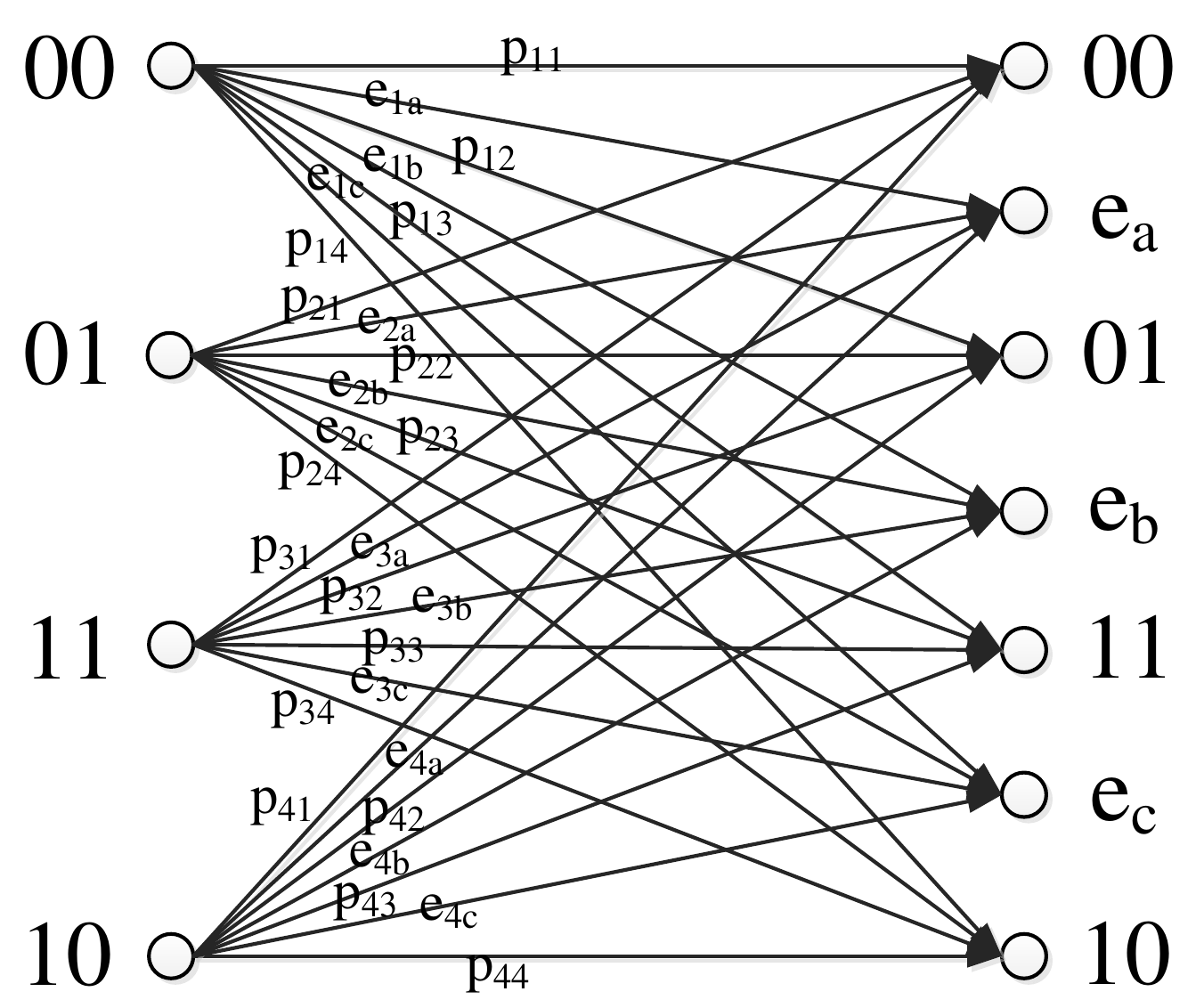}
\caption{Quantization model for 4-MLC with 6 reads.}\label{ch4MLC6reads}
\vspace{0.5in}
\includegraphics[width=0.45\textwidth]{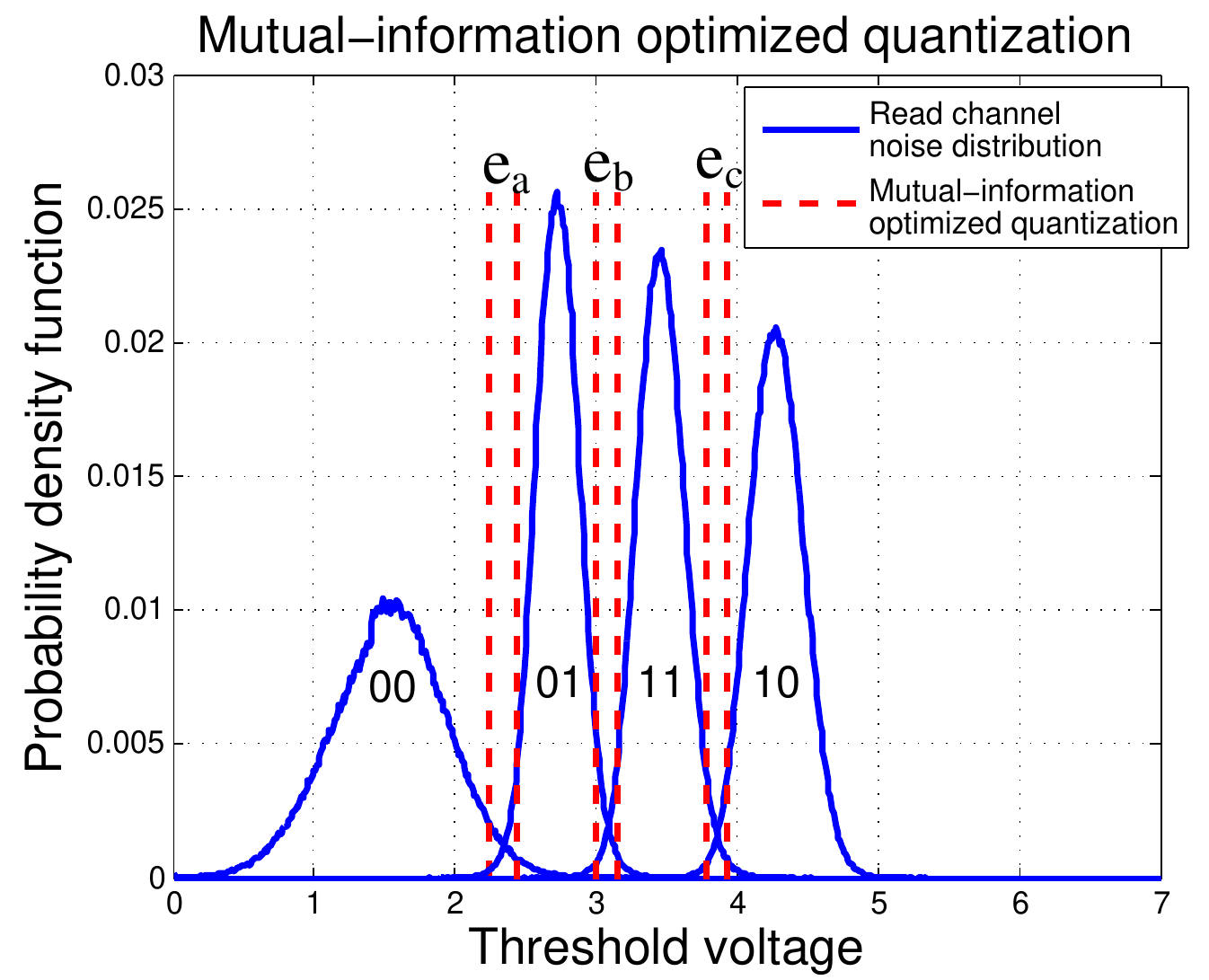}
\caption{Mutual-information optimized quantization for the 6-month data.}\label{fig:pdf}
\end{figure}

If we consider three additional word-line voltages (for a total of six), the threshold voltage can be quantized to seven distinct values as shown in Figure~\ref{quant4MLC6reads}.  The resulting DMC with four inputs and seven outputs is given in Figure~\ref{ch4MLC6reads}. Since the channel is symmetric, the crossover probabilities for the channel model are symmetric in the upper and lower half of the figure, i.e., $ p_{11}=p_{44} $, $ e_{1a}=e_{4c} $, $ p_{12}=p_{43} $, etc.

The qualitative behavior of MMI and required BER as the number of reads is increased in the MLC case is essentially the same as was shown in Figs. \ref{mutual_slc} and \ref{fig:maxerror_vs_Nreads} for the SLC case. 

Similar to the SLC analysis, the MI between the input and output can be calculated as
\begin{align}
I(X;Y)=& H(Y)-H(Y|X)\notag\\
=& H\left(\frac{p_{11}+p_{21}+p_{24}+p_{14}}{4},\frac{p_{12}+p_{22}+p_{23}+p_{13}}{4}, \right. \notag\\
&\frac{p_{13}+p_{23}+p_{22}+p_{12}}{4},\frac{p_{14}+p_{24}+p_{21}+p_{11}}{4},\notag\\
&\frac{e_{1a}+e_{2a}+e_{2c}+e_{1c}}{4},\frac{e_{1b}+e_{2b}+e_{2b}+e_{1b}}{4},\notag\\
&\left. \frac{e_{1c}+e_{2c}+e_{2a}+e_{1a}}{4}\right)\notag\\
&-\frac{1}{2}H(p_{11},p_{12},p_{13},p_{14},e_{1a},e_{1b},e_{1c})\notag\\
&-\frac{1}{2}H(p_{21},p_{22},p_{23},p_{24},e_{2a},e_{2b},e_{2c}),\label{equ:mu6reads}
\end{align}
where all of the crossover probabilities can be calculated in the same manner as with SLC. Thus, in order to choose the optimal quantization levels $ q_1,q_2$, and $q_3 $ for a fixed SNR, we maximize the MI given in equation \eqref{equ:mu6reads}.

\subsection{Retention Model}\label{sec:retention}
We can extend the MMI analysis of Section~\ref{sec:gaussian} to any model for the flash memory read channel.  As an example, consider again the 4-level 6-read MLC as a 4-input 7-output discrete channel as shown in Fig.~\ref{ch4MLC6reads}.  However, instead of assuming Gaussian noise distributions as in Section~\ref{sec:gaussian}, this section numerically computes the probability transition matrix using the retention noise model of \cite{DongUSENIX2011} for a six-month retention period. Fig.~\ref{fig:pdf} shows the {four conditional threshold-voltage} probability density functions generated {according to} \cite{DongUSENIX2011} and the resulting {six} MMI word-line voltages that maximize MMI for this noise model.  {While the conditional noise for each transmitted (or written) threshold voltage is similar to that of a Gaussian, the variance of the conditional distributions varies greatly across the four possible threshold voltages.  Note that the lowest threshold voltage has by far the largest variance.}

Since the retention noise model is not symmetric, we need to numerically compute all the probabilities in Fig.~\ref{ch4MLC6reads} and calculate the MI between the input and output as shown in  \eqref{equ:mu6reads_asym}.  The MI in \eqref{equ:mu6reads_asym} is in general not a quasi-concave function in terms of the word-line voltages $ q_1,q_2,...,q_6 $, although it is quasi-concave for the simple model of two symmetric Gaussians with symmetric word-line voltages studied in \cite{JWANGGLOBECOM11}. Since \eqref{equ:mu6reads_asym} is a continuous and smooth function and locally quasi-concave in the range of our interest, we can numerically compute the MMI quantization levels with a careful use of bisection search.

\begin{figure}
\begin{align}
I(X;Y)=& H(Y)-H(Y|X)\notag\\
=& H\left(\frac{p_{11}+p_{21}+p_{31}+p_{41}}{4},\frac{p_{12}+p_{22}+p_{32}+p_{42}}{4}, \right. \notag\\
&\frac{p_{13}+p_{23}+p_{33}+p_{43}}{4},\frac{p_{14}+p_{24}+p_{34}+p_{44}}{4},\notag\\
&\frac{e_{1a}+e_{2a}+e_{3a}+e_{4a}}{4},\frac{e_{1b}+e_{2b}+e_{3b}+e_{4b}}{4},\notag\\
&\left. \frac{e_{1c}+e_{2c}+e_{3c}+e_{4c}}{4}\right)\notag\\
&-\frac{1}{4}H(p_{11},p_{12},p_{13},p_{14},e_{1a},e_{1b},e_{1c})\notag\\
&-\frac{1}{4}H(p_{21},p_{22},p_{23},p_{24},e_{2a},e_{2b},e_{2c})\notag\\
&-\frac{1}{4}H(p_{31},p_{32},p_{33},p_{34},e_{3a},e_{3b},e_{3c})\notag\\
&-\frac{1}{4}H(p_{41},p_{42},p_{43},p_{44},e_{4a},e_{4b},e_{4c}).\label{equ:mu6reads_asym}
\end{align}
\vspace{-.2in}
\end{figure}

\subsection{The Constant-Ratio Method}\label{sec:constant-ratio}

In \cite{DongTCAS2011},  the constant-ratio (CR) method selects additional word-line voltages to the left and right of each hard-decision word-line voltage using a heuristic that constrains the word-line voltages so that the largest and second-largest conditional noise pdfs have a specified constant ratio $R$.  This method leaves the specification of $R$ to empirical simulation.  The CR method can be viewed as a constraint that can be applied to MMI optimization in order to reduce the search space.   The CR method can also simplify optimization by yielding a convex or quasi-convex problem.

\begin{figure}
\centering
\includegraphics[width=0.49 \textwidth]{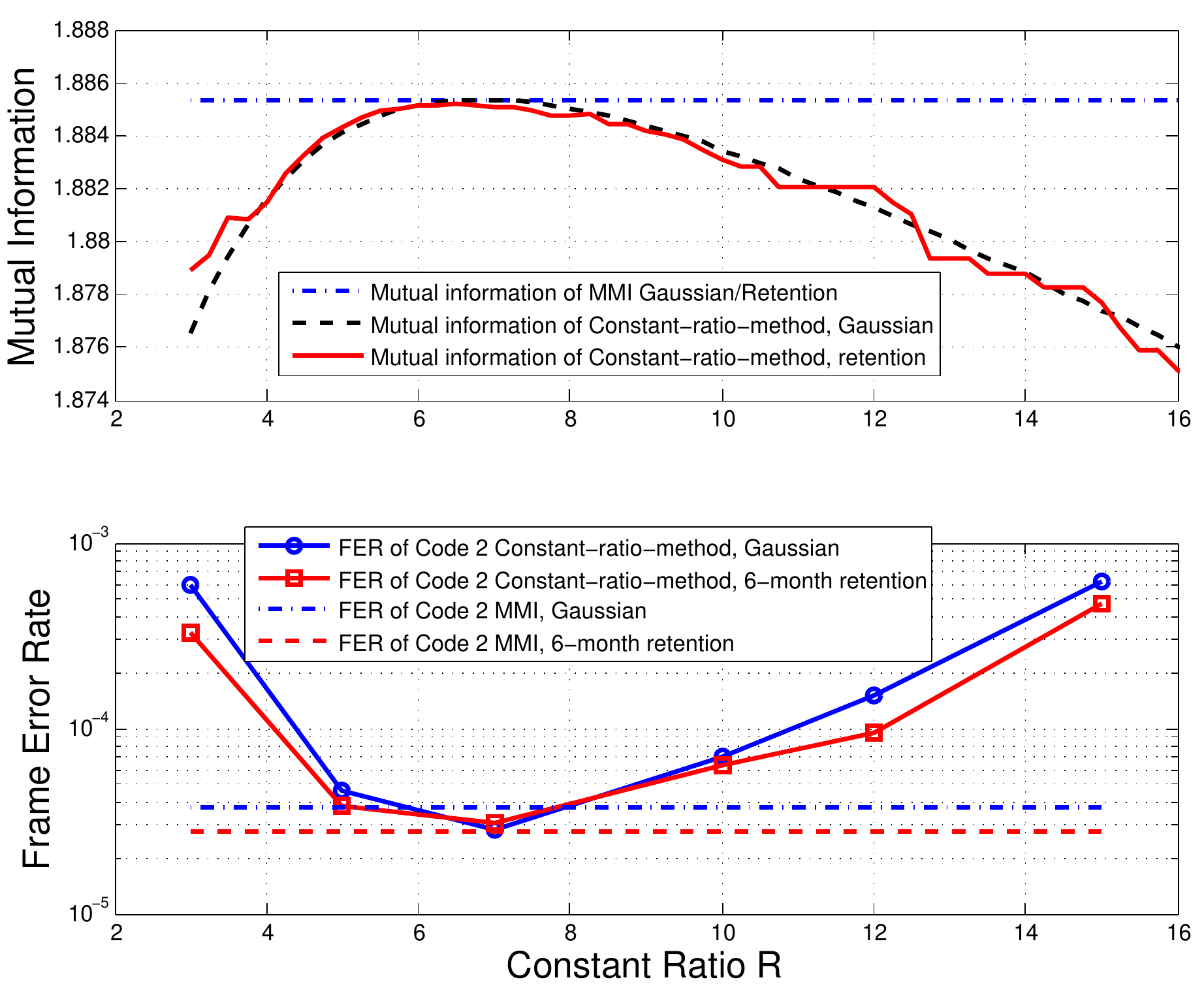}
\caption{Mutual information and frame error rate  separately plotted as functions of the constant-ratio value $R$ for six quantization thresholds (seven levels).  Curves are shown for both the 4-PAM Gaussian model with SNR~$= 13.76$ dB and the retention model of \cite{DongUSENIX2011} for 6 months.  These two models both have an MMI of 1.885 bits shown as a dashed line in the mutual information plot.  The frame error rate plots are for LDPC Code 2 described in Section \ref{results} below.  The two models had slightly different frame error rates with MMI quantization, $3.78 \times 10^{-5}$ for the 4-PAM Gaussian model and $2.8 \times 10^{-5}$ for the retention model shown as dashed lines in the frame error rate plot.}\label{fig:MIvsR}
\vspace{.2in}
\includegraphics[width=0.4\textwidth]{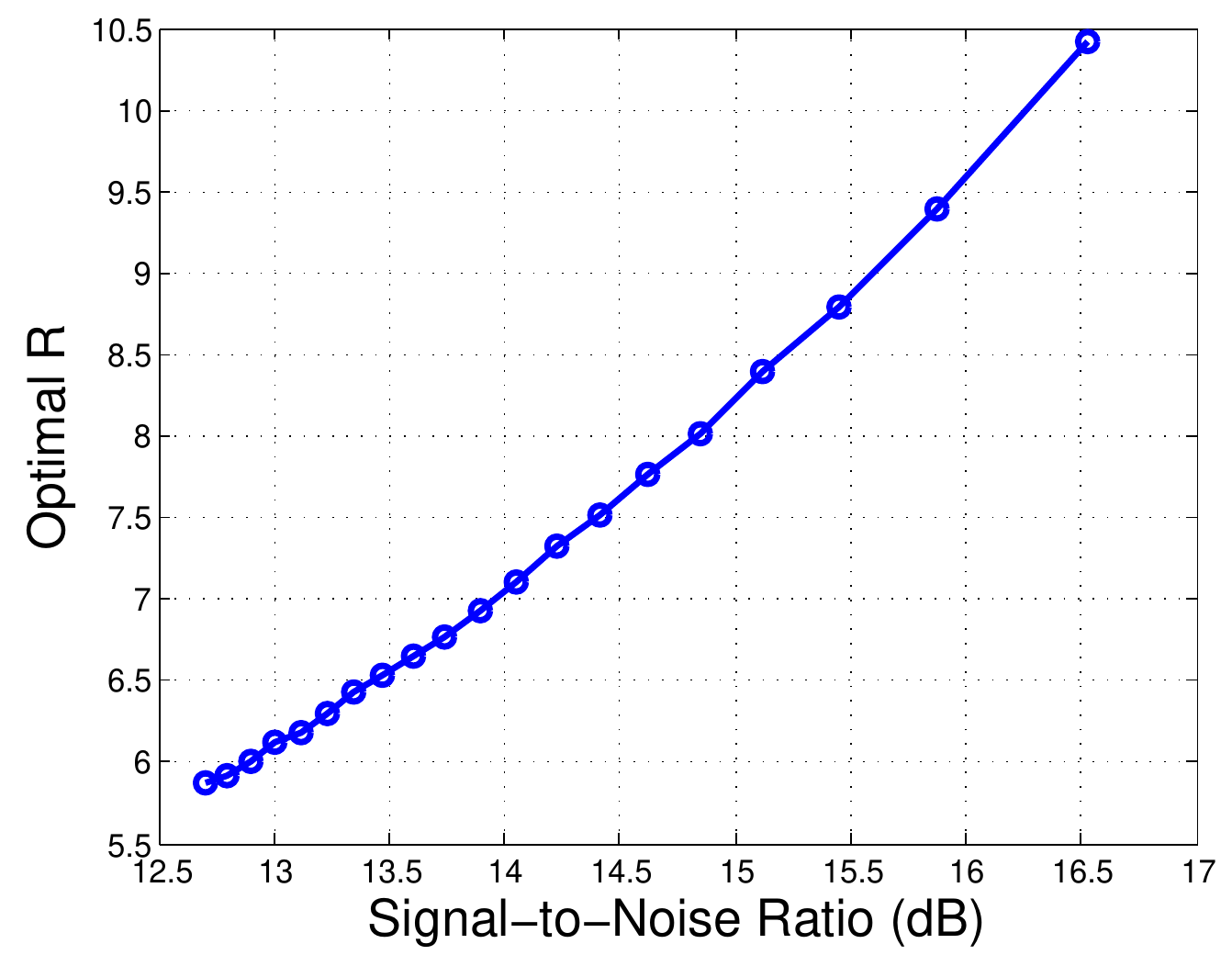}
\caption{Optimal $R$ vs. SNR for four-level MLC with threshold voltages modeled as Gaussians with the same variance.}\label{fig:BestR}
\vspace{-.2in}
\end{figure}

Fig.~\ref{fig:MIvsR} shows MI as a function of $R$ for four-level MLC with six quantization thresholds (seven quantization levels) for both the simple 4-PAM Gaussian model and the more realistic retention model of \cite{DongUSENIX2011}.   The Gaussian and retention channels were selected so that they have an identical MMI for six-read (seven-level) unconstrained MMI optimization.  For both models the CR method with the MI-maximizing $R$ provides essentially the same MI as obtained by the unconstrained MMI optimization.  Furthermore, it is striking how similar the MMI vs. $R$ behavior is for the two different channel models.   For the Gaussian model, MI is a concave function of $R$.  The curve of MI vs. $R$ for the retention model closely follows the Gaussian model curve, but is not a strictly concave function because of variations in the numerical model.  

The MMI approach is a way to select quantization levels in the hope of optimizing frame-error-rate (FER) performance.  Fig.~\ref{fig:MIvsR} shows the FER performance as a function of $R$ for both the Gaussian model and the retention model for LDPC Code 2 described in Section \ref{results} below.  The value of $R$ that provides the maximum MI also delivers the lowest FER as a function of $R$, thus confirming the essential hypothesis of the MMI approach.  Also, the constraint to a constant-ratio does not appear to adversely affect FER since the  lowest FER as a function or $R$ is essentially the same as (and in the Gaussian case even slightly better than) the FER achieved by unconstrained MMI quantization.  The range of MI in Fig.~\ref{fig:MIvsR} is small (approximately 0.01 bits), but this variation in MI corresponds to more than an order of magnitude of difference in FER performance.  

Fig.~\ref{fig:MIvsR} focused on two channels with the same MMI and we saw that the optimal value of $R$ was 7 for those channels.  Fig. \ref{fig:BestR} shows how the optimal value of $R$ (the value that maximizes the mutual information) varies with SNR for the case of four-level MLC with threshold voltages modeled as Gaussian distributions all with the same variance as in Fig. \ref{quant4MLC6reads}.




\section{Simulation Results}\label{results}


In this section we demonstrate the benefits of LDPC decoding using soft information provided through multiple reads.  A rate-0.9021 BCH code with block length $n=9152$ and dimension $k=8256$ provides the baseline for comparison. For our simulations, we considered three rate-0.9021 irregular LDPC codes with block length $ n=9118 $ and dimension $ k=8225 $.  Two of the LDPC codes studied (Codes 1 and 2) feature distinct degree distributions with maximum variable degree 19.  Code 3 has a larger maximum variable degree of 24.  The LDPC matrices\footnote{The complete LDPC code parity-check matrices are available at the CSL website \url{http://www.ee.ucla.edu/~csl/files/publications.html#COD}.}  were constructed according to their respective degree distributions using the ACE algorithm \cite{TianTCOM04}, and the stopping-set check algorithm \cite{RamamoorthyICC04}.   All of the simulations were performed using a sequential belief propagation decoder.   The frame sizes are the block lengths, $k=8256$ for BCH and $k=8225$ for LDPC.  

\subsection{LDPC code design in a quantized setting}\label{sec:LDPC_quantization}
LDPC code design is not the focus of this paper, but it is important to note that code design for a quantized system differs from code design for an AWGN channel where full precision soft information is available.   For Code 1, the degree distribution is the usual optimal degree distribution for AWGN \cite{RichardsonDes} (i.e. the code with the best density-evolution threshold).  However, the channel produced by hard decoding is the binary symmetric channel rather than the Gaussian channel.  

\begin{figure}[t]
\centering
\includegraphics[width=0.35\textwidth]{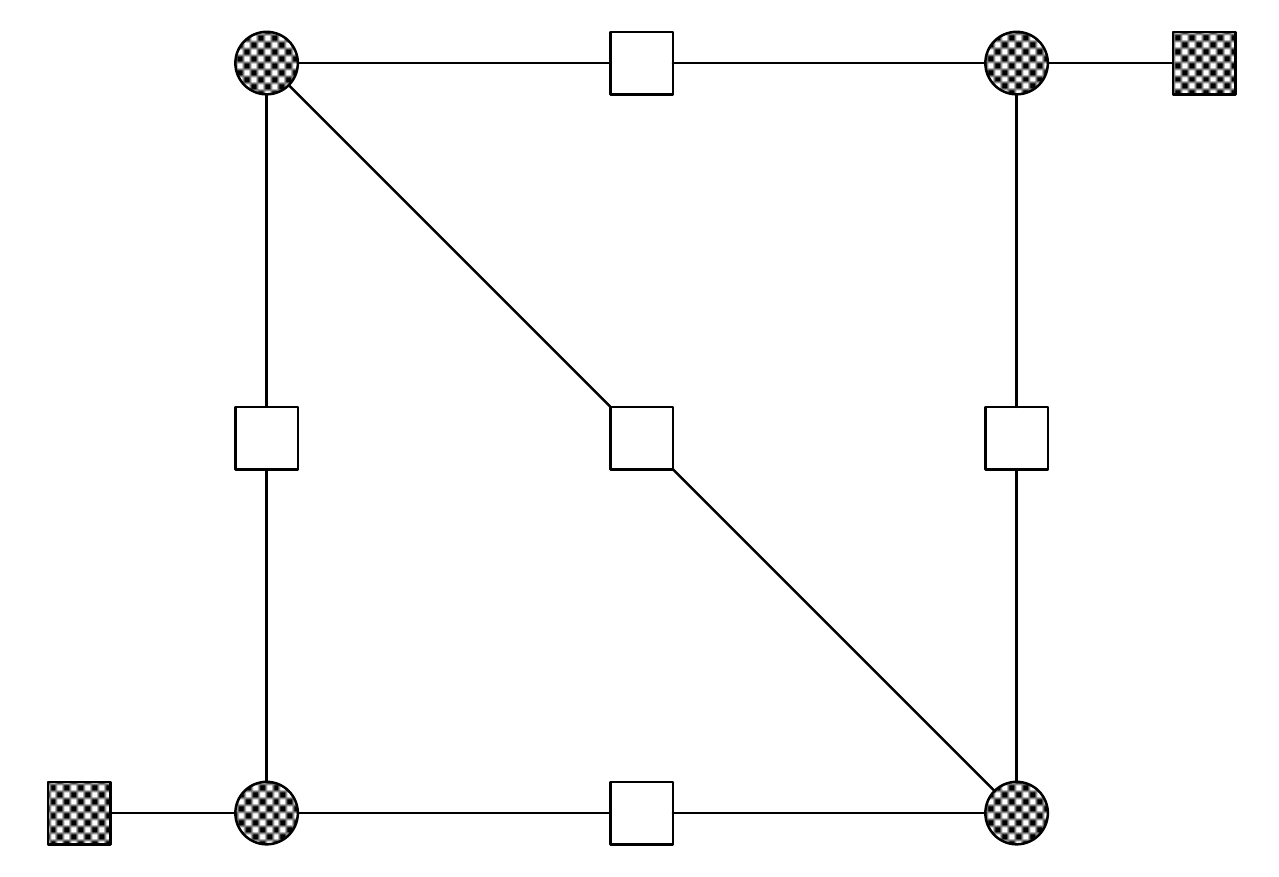}
\caption{A (4,2) absorbing set.  Variable nodes are shown as black circles.  Satisfied check nodes are shown as white squares.   Unsatisfied check nodes are shown as black squares.  Note that each of the four variable nodes has degree three.  This absorbing set is avoided by precluding degree-3 nodes. }\label{fig:abs42}
\end{figure}

Hard decoding suffers from errors induced by small absorbing sets \cite{dolecekIT10,WangIT} such as the $(4,2)$, $(5,1)$, and $(5,2) $ absorbing sets in Code 1.  Fig. \ref{fig:abs42} shows the $(4,2)$ absorbing set for illustration.  As shown in Fig. \ref{fig:abs42} for the (4,2) absorbing set, the $(4,2)$, $(5,1)$, and  $(5,2) $ absorbing sets can all be avoided by precluding degree-three variable nodes.  All the degree-3 variable nodes are increased to have degree 4 to produce Code 2.  With these absorbing sets removed, Code 2 significantly outperforms Code 1 under hard decoding as will be shown in Figs. \ref{sim:mlc4} and \ref{fig:simretention_code2}. 

We also simulated another code (Code 3) that retained the original AWGN-optimized degree distribution but increased the maximum variable node degree from 19 to 24 in an attempt to improve performance in hard decoding.  Code 3 has the best density-evolution threshold in AWGN of these three codes, but Code 2, with the worst AWGN threshold of the three codes, provides the best performance under hard decoding (as we will see below) because its degree distribution precludes the problematic absorbing sets. This demonstrates that a superior AWGN threshold does not necessarily imply superior performance under hard decoding.  

However, when simulated in AWGN with full resolution {\em soft} decoding, Codes 1 and 3 (which were designed for AWGN with full-resolution soft decoding) both perform better than Code 2 as shown in Figs. \ref{sim:slc} and \ref{sim:mlc4} below.  Thus these codes present a trade-off between performance under hard-decoding (or limited precision) and performance under full-precision soft decoding.  

For typical Flash applications, hard decoding and limited-precision decoding would be employed because of complexity and latency constraints.  Hence,  a code such as Code 2 is the clear choice.  However,  an interesting area of future research is the development of codes that are ``universal'' across precision variations.  It would be useful to design a code that can perform well over a large range of precisions or to show that such universal performance is not possible.

\subsection{Frame error rate improvement with increased precision}

\begin{figure}
\centering
\includegraphics[width=0.45\textwidth]{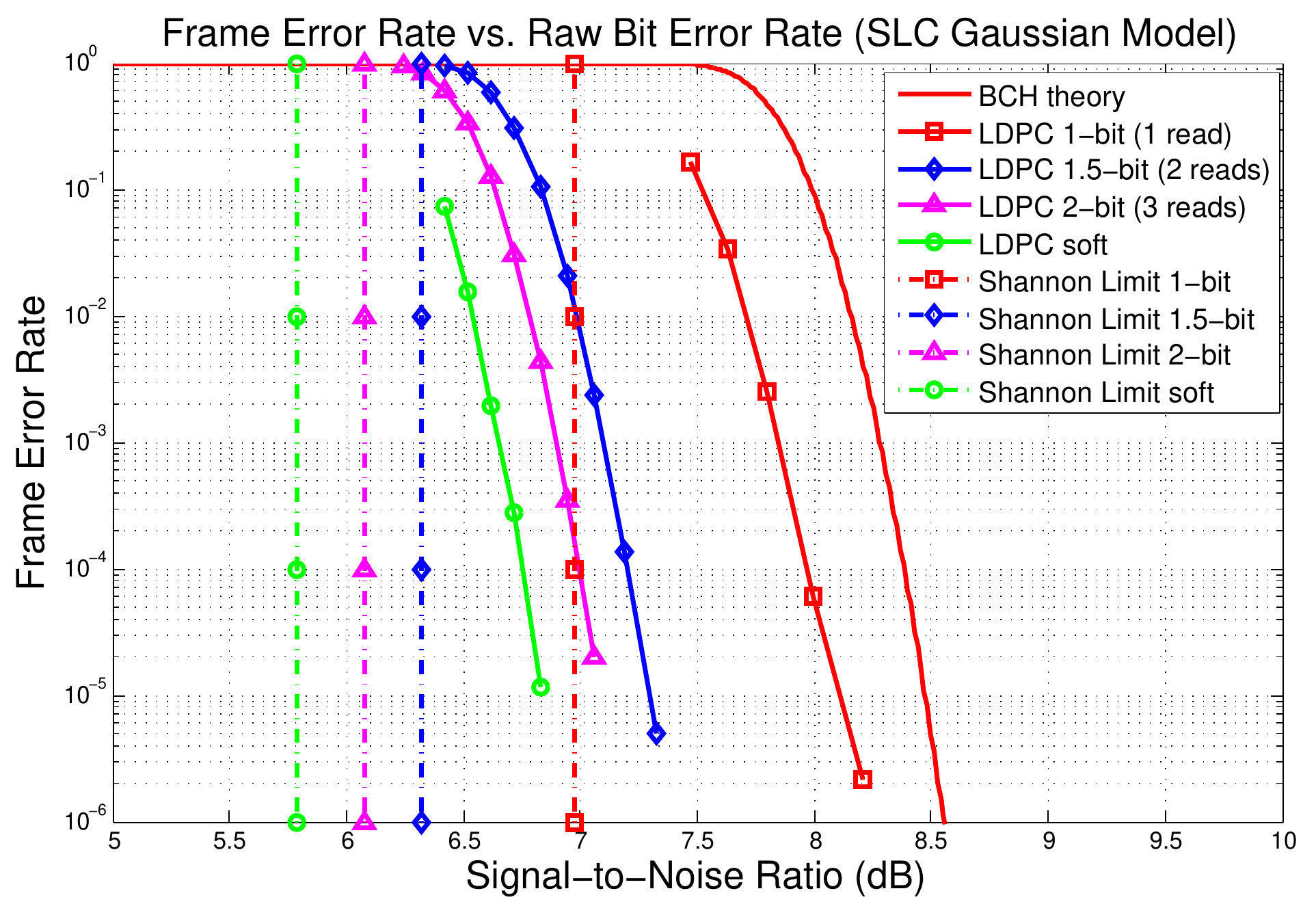}
\caption{FER vs. SNR simulation results using the Gaussian channel model for SLC comparing LDPC Code 2 with varying levels of soft information and a BCH code with hard decoding.  Both codes have rate 0.9021.  The BCH theory curve and the LDPC 1-read curve correspond to hard decoding.}\label{sim:slc_snr}
\end{figure}

This subsection looks at how LDPC performance improves as more soft information is made available to the decoder.  For readers familiar with the standard FER curves presented in the communications literature,  Fig. \ref{sim:slc_snr} plots FER versus the traditional signal-to-noise ratio (SNR) $\frac{E_s}{N_0/2} $ for LDPC Code 2 and the BCH code both with rate 0.9021.   The performance of these codes is studied on the Gaussian model of the  SLC Flash memory cell.   This plot illustrates that each additional read improves the FER performance of the LDPC code, but the performance improvement is diminishing.

\begin{figure}
\centering
\includegraphics[width=0.5\textwidth]{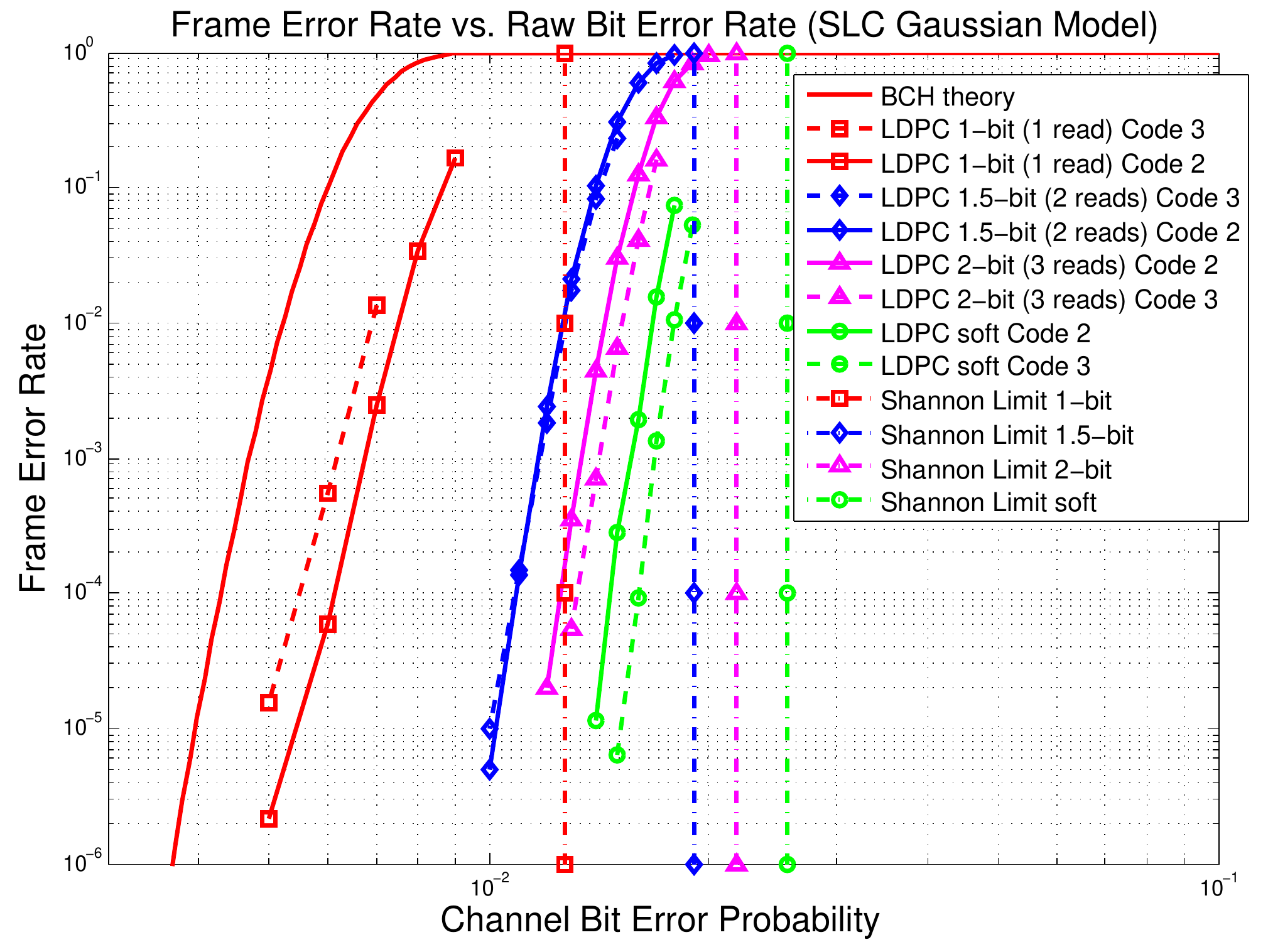}
\caption{FER vs. channel bit error probability simulation results using the Gaussian channel model for SLC comparing LDPC Codes 2 and 3 with varying levels of soft information and a BCH code with hard decoding.  All codes have rate 0.9021.  The BCH theory curve and the LDPC 1-read curve correspond to hard decoding.}\label{sim:slc}
\vspace{0.2in}
\includegraphics[width=0.45\textwidth]{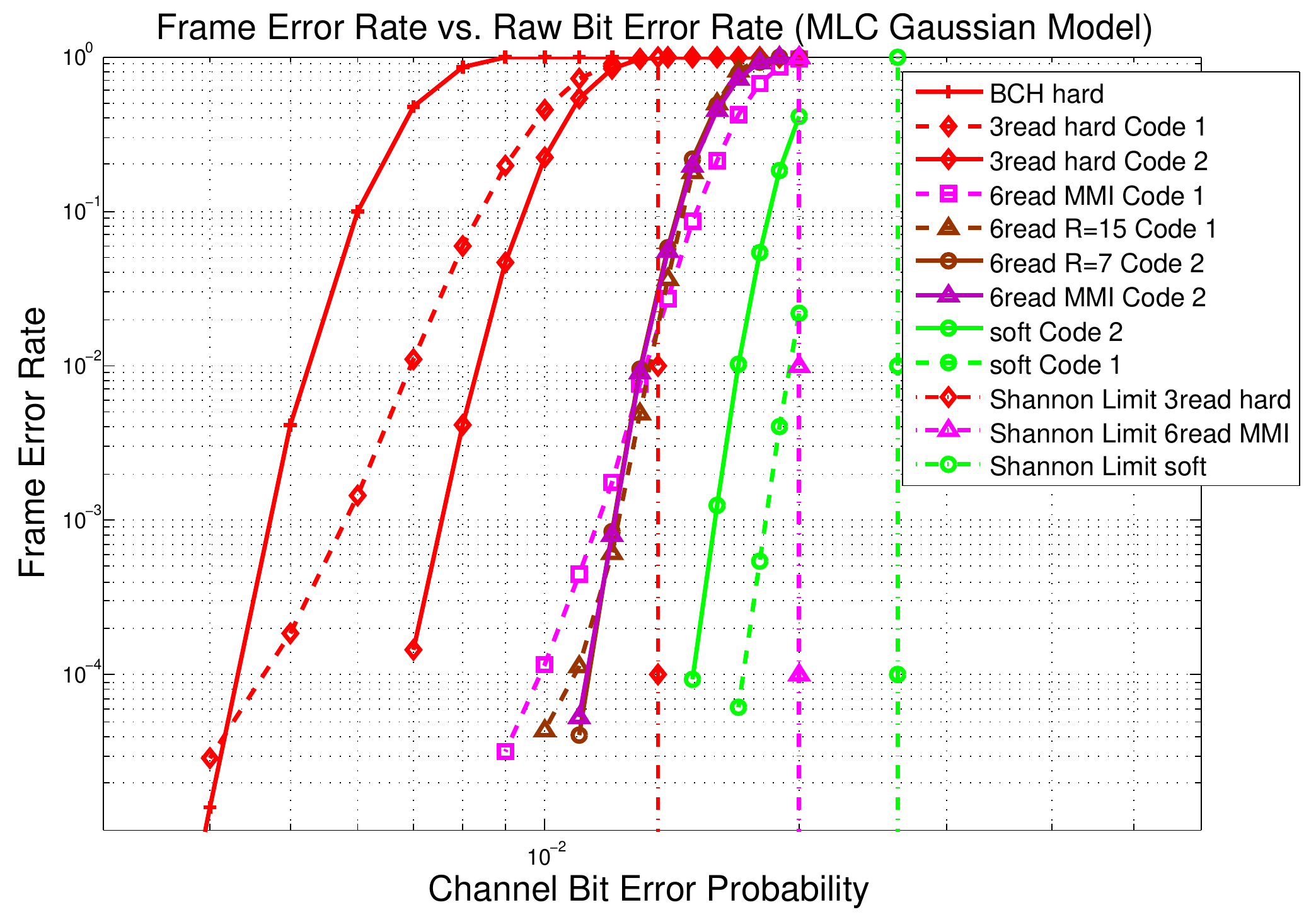}
\caption{FER vs. channel bit error probability simulation results using the Gaussian channel model for 4-level MLC comparing LDPC Codes 1 and 2 with varying levels of soft information and a BCH code with hard decoding.  All codes have rate 0.9021. The BCH hard curve and the LDPC 3-read curves correspond to hard decoding.}\label{sim:mlc4}
\end{figure}

Plots showing  FER vs. the channel bit error probability, which is the rate of hard decoding errors on the read channel, are the most common way of presenting performance of Flash memory systems.  Fig. \ref{sim:slc} plots the data of Fig. \ref{sim:slc_snr} in this way.   Each SNR of Fig. \ref{sim:slc_snr} has a corresponding channel bit error probability in Fig.  \ref{sim:slc}.   Whether considered from the perspective of SNR in  Fig. \ref{sim:slc_snr} or channel bit error probability in Fig.  \ref{sim:slc}, the soft information provided by two reads recovers more than half of the gap between hard decoding (one read) and full soft-precision decoding. 

Fig. \ref{sim:mlc4} provides a similar plot for the Gaussian model of four-level MLC.  With four levels, three reads are required for hard decoding.  Similar to the discussion above, using six reads recovers more than half of the gap between hard decoding (three reads) and full soft-precision decoding. 

The principle of closing the gap by more than half with the first additional read (or first three additional reads in the MLC case) and diminishing returns for subsequent reads can also be observed by examining the vertical dashed lines showing the Shannon limits corresponding to varying levels of soft information in Figs.~\ref{sim:slc_snr}-\ref{sim:mlc4}. 

\subsection{Performance of BCH and the three LDPC designs}
This subsection looks more closely at performance differences between the various codes studied.
Since the BCH decoder is limited to using hard decisions from the comparator, we first compare LDPC code performance to the BCH code using only hard decisions in order to make a fair baseline comparison.   The BCH and LDPC hard-decoding curves in Figs. \ref{sim:slc} and \ref{sim:mlc4} (the red curves) show that the LDPC codes outperform the BCH code in this range of frame error rates even under hard decoding, with the notable exception of Code 1 in Fig. \ref{sim:mlc4} whose error floor crosses the BCH curve at an FER of $4 \times 10^{-5}$. For reference, the red dashed vertical line gives the Shannon limit for operating at rate 0.9021 on this channel with hard decoding.

Note that as discussed above in Section \ref{sec:LDPC_quantization}, for hard decoding Code 2 outperforms both Code 3 in Fig.  \ref{sim:slc} and Code 1 in Fig. \ref{sim:mlc4}.  As more and more precision becomes available to the decoder, the performance of Codes 1 and 3 improves relative to Code 2.  

In Fig.  \ref{sim:slc} the FER curves for Codes 2 and 3 are identical when two reads are performed, and Code 3 has a small performance advantage over Code 2 when three reads are performed.   Note that Code 3 has a larger maximum variable node degree than Code 2 (24 vs. 19). 

In Fig.  \ref{sim:mlc4} Code 2 still has an advantage over Code 1 when six reads are performed.  Ultimately, when full-resolution soft information is available, the Code 2 performance  is inferior to that of Codes 1 and 3.

Of course the BCH code would also benefit from the use of soft information.   However, soft decoding of BCH codes is not commonly performed because of the associated complexity.   BCH decoding utilizing soft information (such as erasures) is outside the scope of this paper.

\begin{figure}
\centering
\includegraphics[width=0.45\textwidth]{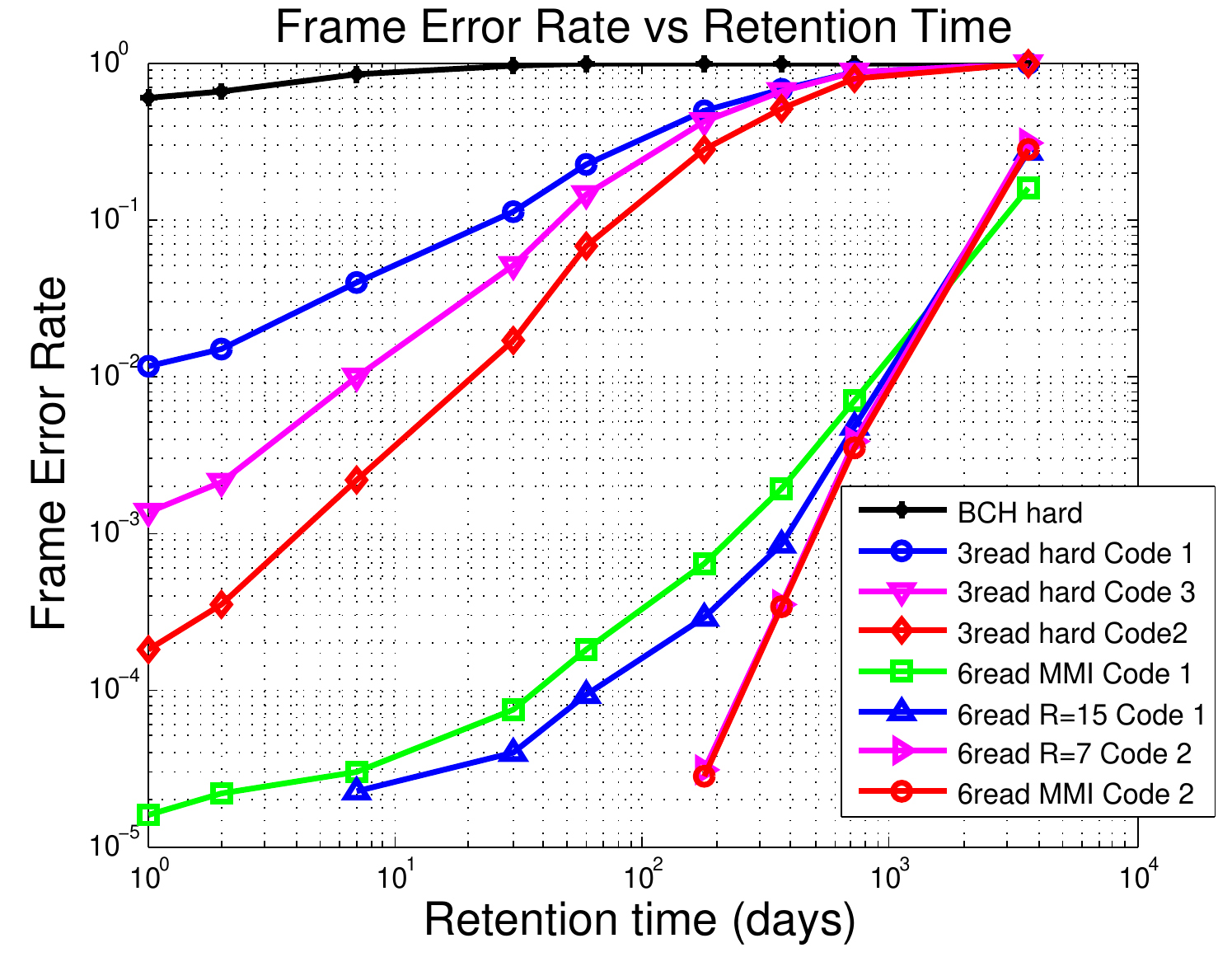}
\caption{FER vs. retention time simulation results using the retention model of \cite{DongUSENIX2011} for 4-level MLC.  All codes have rate 0.9021. Hard decoding results are shown for the BCH code and LDPC Codes 1, 2, and 3.   FER performance for higher-precision decoding using six reads is shown for LDPC Codes 1 and 2 using both unconstrained MMI quantization and MMI quantization with the constant-ratio constraint  with the $R$ value that maximizes MI for that LDPC code.  }\label{fig:simretention_code2}
\end{figure}

Now we examine code performance using the retention model of \cite{DongUSENIX2011}.  Fig.~\ref{fig:simretention_code2} shows frame error rate (FER) plotted versus retention time for Codes 1, 2, and 3 with three reads and for Codes 1 and 2 for six reads.  

The  three-read quantization whose performance is shown in Fig. \ref{fig:simretention_code2} is  standard hard decoding for four-level MLC.   We note that in principle, since the retention model is not symmetric, some gain can be achieved by adjusting the thresholds using MMI even in the three-read case.  However, we found those gains to be insignificant in our simulations. 

As mentioned in Section \ref{sec:LDPC_quantization}, Code 2 outperforms both Code 3  and Code 1 under hard decoding in Fig.  \ref{fig:simretention_code2}.  
For six reads, Code 2 still outperforms  Code 1.  However, Code 3 with six reads (not shown) performs slightly better than Code 2 achieving an FER of $2 \times  10^{-4}$ at a retention time of 400 days with $R=10$ CR-method quantization.  Note again that Code 3 has a larger maximum variable node degree than Code 2. 

\subsection{Unconstrained-MMI, MMI-CR, and ``best $R$'' quantization}

This subsection examines the simulation results of Figs. \ref{sim:mlc4}  and \ref{fig:simretention_code2} to compare unconstrained  MMI quantization, MMI quantization constrained by the CR method, and the CR method with the value of $R$ that empirically provides the lowest FER.

A ``good code'' that performs close to the theoretical limits (i.e. the mutual information) will perform best when the theoretical limit is at its best (i.e. when the mutual information is maximized).  Hence it is not too surprising that MMI quantization brings out the best performance from Code 2.   In both Fig. \ref{sim:mlc4}  and Fig. \ref{fig:simretention_code2}, the Code-2 FER curves for unconstrained-MMI quantization and for $R=7$ and indistinguishable.  Note that $R=7$ is both the MMI value of $R$ and the value of $R$ that empirically minimizes FER for Code 2.

However, if a code is not a ``good code'' and is operating relatively far from the theoretical limits, this argument does not hold.  Thus, a code with relatively poor performance might perform slightly better with a quantization that does not maximize the mutual information.  Indeed, the best FER performance for Code 1 for six reads with CR quantization is with $R=15$ both for the Gaussian model of Fig.  \ref{sim:mlc4}  and the retention model of Fig. \ref{fig:simretention_code2}.  Note from Fig. \ref{fig:MIvsR} that $R=15$ provides a smaller mutual information than $R=7$.  Notice that for Code 1 with six reads the unconstrained MMI quantization performs slightly worse that the $R=15$ quantization.  Thus we can see that for a weaker code, the MMI approach may not provide the best possible quantization in terms of FER.  However, this situation may well be interpreted as an indicator that it may be worth exploring further code design to improve the code.

To avoid crowding the plot of Fig. \ref{fig:simretention_code2}, the performance of Code 3 with six reads is not shown.  However, the lowest FER for Code 3 with six reads is obtained with $R=10$ and is slightly better than Code 2 achieving an FER of $2 \times  10^{-4}$ at a retention time of 400 days.  This performance is also slightly better than the FER of the unconstrained-MMI selection of word-line voltages for Code 3, which is indistinguishable from the FER performance of Code 2.   Even though it slightly outperforms Code 2 because of its larger maximum variable node degree, the fact that the MMI value of $R$ does not minimize the FER of Code 3 is an indication that it can be further improved for this quantized channel.

\section{Conclusion}\label{conclusion}
This paper explores the benefit of using soft information in an LDPC decoder for NAND flash memory. Using a small amount of soft information improves the performance of LDPC codes significantly and demonstrates a clear performance advantage over conventional BCH codes. 

In order to maximize the performance benefit of the soft information, we present an approach for optimizing word-line-voltage selection so that the resulting quantization maximizes the mutual information between the input and output of the equivalent read channel. This method can be applied to any channel model.  Constraining the quantization using the constant-ratio method provides a significant simplification with no noticeable loss in performance.   

Our simulation results suggest that if the LDPC code is well designed, the quantization that maximizes the mutual information will also minimize the frame error rate.  However, we also saw that care must be taken to design the code to perform well in the quantized channel.  An LDPC code designed for a full-precision Gaussian channel may perform poorly in the quantized setting.  




\bibliographystyle{unsrt}	

\bibliography{myrefs}		

\end{document}